\documentclass[twocolumn,trackchanges]{aastex701}
\usepackage{amsmath}
\usepackage{graphicx}
\usepackage{indentfirst}
\usepackage{float}
\usepackage{subcaption}
\usepackage{caption}
\usepackage{multirow}
\usepackage{CJK}
\usepackage{threeparttable}
\usepackage{amsmath}

\renewcommand{\thefootnote}{\fnsymbol{footnote}}

\begin{document}
\begin{CJK*}{UTF8}{gbsn} 

\title{Multi-scale Memory and Regime Shift in the Hyperactive Repeating FRB 20240114A}

\author[0009-0008-6247-0645]{Wen-Long Zhang (张文龙)}
\affil{Purple Mountain Observatory, Chinese Academy of Sciences, Nanjing 210023, China}
\affil{School of Astronomy and Space Sciences, University of Science and Technology of China, Hefei 230026, China}
\email{wlzhang@pmo.ac.cn}

\author[0000-0001-9217-7070]{Sheng-Lun Xie (谢升伦)}
\affiliation{School of Mathematics and Physics, Jinggangshan University, Ji'an, Jiangxi 343009, China}
\affiliation{Institute for Astronomy and Astrophysics, Department of Physics, Jinggangshan University, Ji'an, Jiangxi 343009, China}
\email{xiesl@jgsu.edu.cn}

\author[0000-0003-0162-2488]{Jun-Jie Wei* (魏俊杰)}
\affil{Purple Mountain Observatory, Chinese Academy of Sciences, Nanjing 210023, China}
\affil{School of Astronomy and Space Sciences, University of Science and Technology of China, Hefei 230026, China}
\email[show]{jjwei@pmo.ac.cn}

\author[0000-0002-4304-2759]{Di Xiao (肖笛)}
\affil{Purple Mountain Observatory, Chinese Academy of Sciences, Nanjing 210023, China}
\affil{School of Astronomy and Space Sciences, University of Science and Technology of China, Hefei 230026, China}
\email[]{dxiao@pmo.ac.cn}

\author[0009-0002-3020-9123]{Long-Xuan Zhang (张珑萱)}
\affiliation{School of Physics, Huazhong University of Science and Technology, Wuhan, 430074. China}
\email[]{d202580060@hust.edu.cn}

\author[0000-0003-0672-5646]{Shuang-Xi Yi (仪双喜)}
\affil{School of Physics and Physical Engineering, Qufu Normal University, Qufu, Shandong 273165, China}
\email{yisx2015@qfnu.edu.cn}

\author[0000-0003-4157-7714]{Fa-Yin Wang (王发印)}
\affil{School of Astronomy and Space Science, Nanjing University, Nanjing 210023, China}
\email{fayinwang@nju.edu.cn}

\author[0000-0002-6299-1263]{Xue-Feng Wu* (吴雪峰)}
\affil{Purple Mountain Observatory, Chinese Academy of Sciences, Nanjing 210023, China}
\affil{School of Astronomy and Space Sciences, University of Science and Technology of China, Hefei 230026, China}
\email[show]{xfwu@pmo.ac.cn}

\begin{abstract}
We present a statistical analysis of FRB~20240114A, a hyperactive repeating fast radio burst, based on 11,553 bursts detected by FAST over 214 days. 
Our main findings are fourfold.
(1) On the most active day (MJD~60381, 3,197 bursts in 4.38 hr), event-rate coherence analysis reveals persistent correlated activity extending up to 3600~s, the longest reported for any repeating FRB, showing memory persists even in intense bursting epochs. (2) The waiting-time distribution on this day is well described by three exponentials, whereas the full 214-day sample develops a threshold power-law tail, indicating burst statistics depend on the observational baseline, with long-range correlations emerging only over longer timescales, a hallmark of self-organized criticality.
(3) Rescaled range (R/S) analysis of waiting times reveals a broken power law, with Hurst exponents $H_1=0.63\pm0.02$ (short-lag weak memory) and $H_2=1.04\pm0.02$ (long-lag non-stationary drift). The break corresponds to $\sim$1 hour, consistent with the 3600~s coherence limit. R/S analysis of energies similarly exhibits a break ($H_1=0.60\pm0.01$, $H_2=1.10\pm0.05$) at a different lag, reinforcing that non-stationarity affects both temporal and energetic properties.
(4) Energy distributions exhibit waiting-time-dependent slopes that are consistent with the full and daily samples, and the high-energy cutoff remains constant across waiting-time groups, suggesting that the maximum energy scale is an intrinsic source property.
Together, these results establish a multi-scale memory framework: the source behaves stochastically on short timescales but exhibits systemic non-stationarity over months, providing benchmarks for burst models and highlighting the need for long-term, high-cadence monitoring to capture temporal complexity.
\end{abstract}

\keywords{\uat{Radio transient sources}{2008} -- \uat{Astrostatistics}{1882} -- \uat{Neutron stars}{1108}}

\section{INTRODUCTION} 
Fast Radio Bursts (FRBs) are luminous, millisecond-duration radio pulses whose physical origin remains one of the most intriguing puzzles in modern astrophysics. Since the first discovery nearly two decades ago \citep{2007Sci...318..777L}, the field has expanded rapidly, with thousands of sources now detected, including a growing number of repeating FRBs \citep{2019A&ARv..27....4P,2020Natur.587...45Z,2021SCPMA..6449501X,2023RvMP...95c5005Z}.

A major breakthrough came in 2020, when FRB~20200428 was detected from the Galactic magnetar SGR~J1935+2154, accompanied by a bright X-ray burst \citep{2020Natur.587...54C,2020Natur.587...59B,2021NatAs...5..372R,2021NatAs...5..378L}. This association firmly established magnetars (neutron stars with extreme magnetic fields ($\gtrsim10^{14}$~G)) as progenitors of at least some FRBs \citep{2020ApJ...897L..40D,2020ApJ...898L..55G,2020ApJ...904L...5X,2020MNRAS.499.2319K,Wang2022}. The episodic bursting activity of magnetars is thought to be powered by magnetic reconnection and crustal fractures. By analogy with earthquakes and solar flares, the framework of self-organized criticality (SOC; \citealt{1986JGR....9110412K,1987PhRvL..59..381B}) has been applied to magnetar bursts and FRBs \citep{2017JCAP...03..023W,2019ApJ...882..108W,2020MNRAS.491.1498C,2021ApJ...920L..23Z,2023RAA....23k5013Z,2024ApJ...967..108X}. SOC systems generically exhibit scale-free power-law distributions and long-range temporal correlations, both of which have been observed in various astrophysical burst phenomena \citep{1999PhRvL..83.4662B,2013NatPh...9..465W,2020MNRAS.491.1498C,2022RAA....22f5012Z,2025A&A...693A.290Z, 2026arXiv260817964Z,2023PhRvR...5a3019W}.

A central question is whether FRB burst sequences are intrinsically random (Poisson-like) or possess memory. Some analyses report evidence for clustering and power-law tails consistent with SOC \citep{2023ApJ...949L..33W,2024ApJ...975..188W}, whereas others argue for nearly Poissonian behavior \citep{Sang2024, 2024SciBu..69.1020Z}. These seemingly contradictory findings may reflect a deeper underlying reality: the statistical properties may depend on the observational timescale considered.

FRB~20240114A, discovered by the Canadian Hydrogen Intensity Mapping Experiment (CHIME; \citealt{2026ApJ...997..334S}) and extensively monitored with the Five-hundred-meter Aperture Spherical radio Telescope (FAST), has yielded 11,553 bursts over 214 days \citep{2025arXiv250714707Z,2026ApJ...998..276Z,2026SCPMA..6949512Z}, constituting the largest single-source sample to date. This exceptional dataset provides a unique opportunity to probe memory properties over an unprecedentedly wide dynamic range.

In this paper, we employ a combination of event-rate coherence analysis, waiting-time distributions, rescaled range (R/S) analysis, and energy distributions to systematically investigate the memory properties of FRB~20240114A. Section~\ref{sec:data} describes the data and methods, Section~\ref{sec:results} presents the results, Section~\ref{sec:discussion} discusses the physical implications, and Section~\ref{sec:conclusions} summarizes our conclusions.

\section{DATA AND METHODS}
\label{sec:data}

\subsection{Observational Data}
We use the burst catalog of FRB~20240114A presented by \citet{2025arXiv250714707Z}, which was obtained with FAST. The monitoring campaign spanned MJD 60337--60552, corresponding to 33.86~hr of on-source time, and yielded a total of 11,553 independent bursts. The isotropic-equivalent energy of each burst is computed as
\begin{equation}
E = 10^{39}\,\mathrm{erg}\, \frac{4\pi}{1+z}\left(\frac{D_L}{10^{28}\,\mathrm{cm}}\right)^2 \left(\frac{F_\nu}{\mathrm{Jy\,ms}}\right)\left(\frac{\Delta\nu}{\mathrm{GHz}}\right),
\end{equation}
where $D_L=633.9$~Mpc for $z=0.1306$ \citep{2025ApJ...980L..24C}, $F_\nu$ is the measured fluence, and $\Delta\nu$ is the observing bandwidth. The average burst rate is $\sim341~\mathrm{hr^{-1}}$, with a peak of $729~\mathrm{hr^{-1}}$ on MJD~60381.

\begin{figure*}[http!]
\centering
\includegraphics[width=0.7\textwidth]{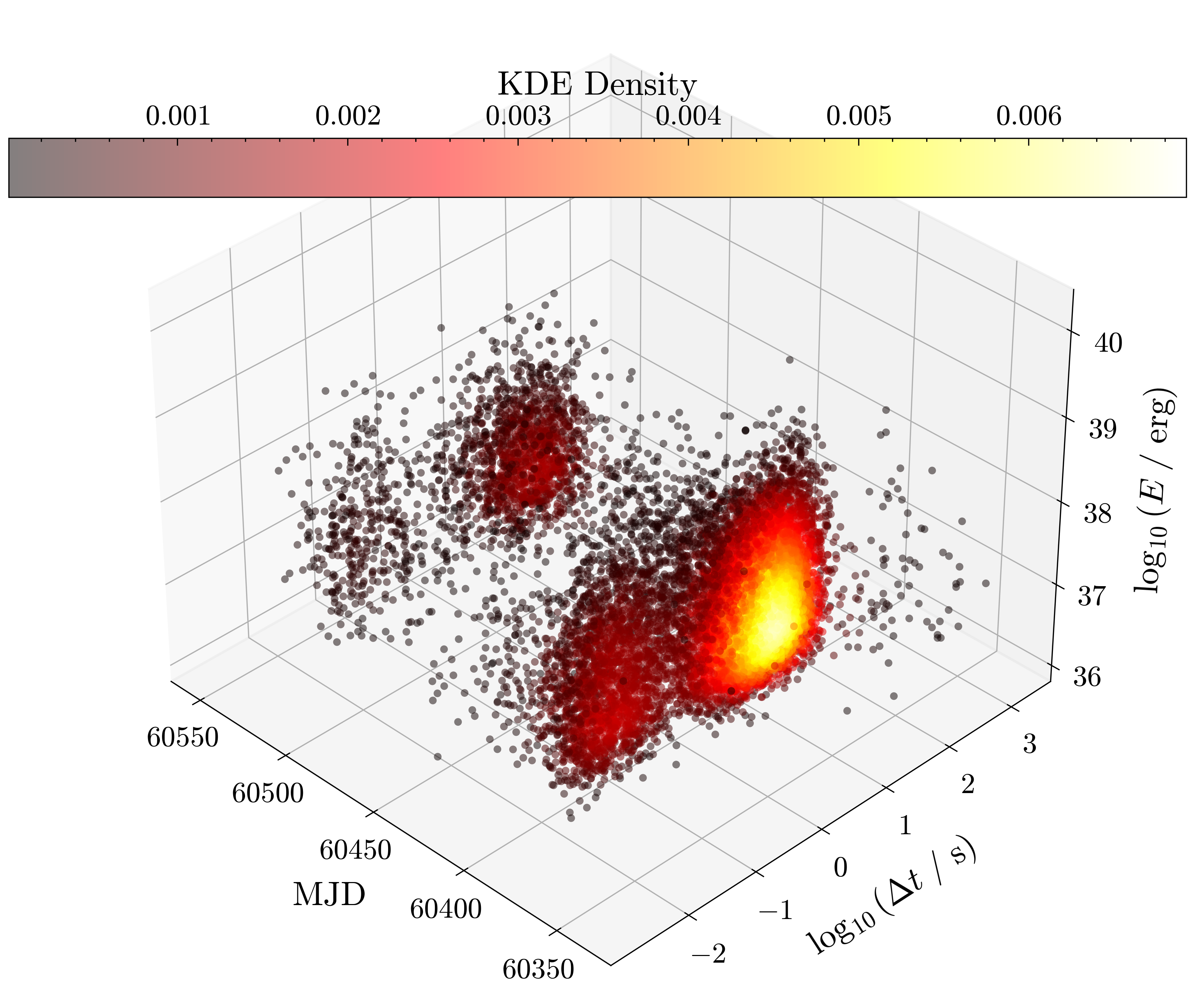}\\
\includegraphics[width=\textwidth]{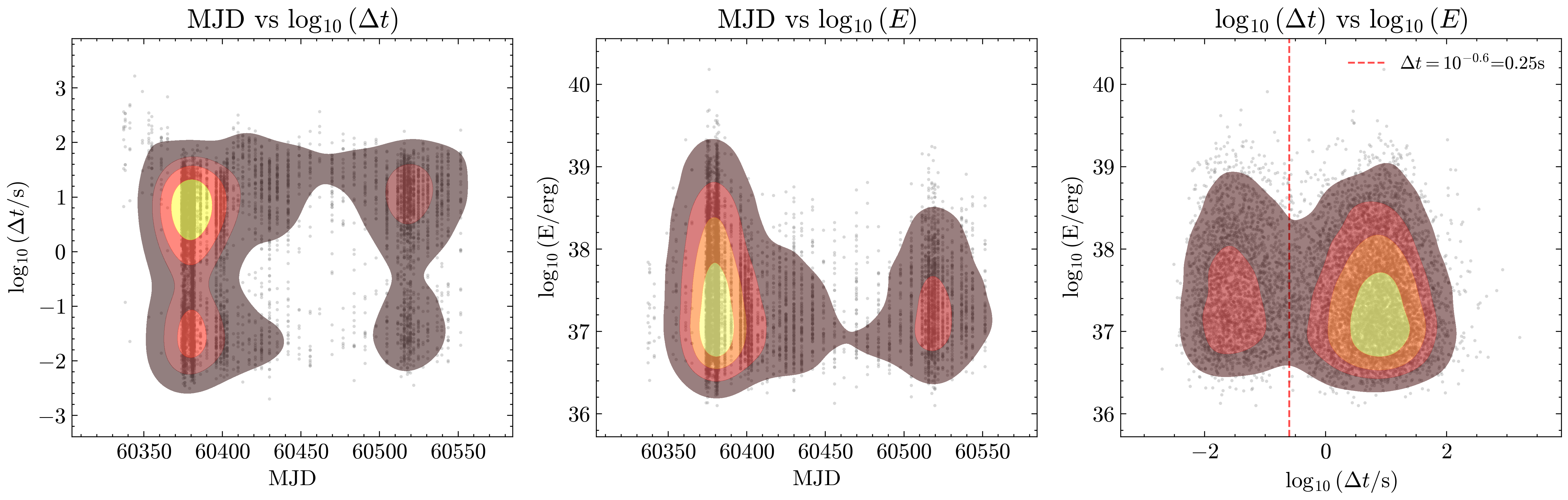}\\
\caption{Three-dimensional KDE of burst energy vs. waiting time vs. MJD (top) and 2D marginal projections (bottom). Burst activity is predominantly concentrated within a few active epochs, with MJD~60381 being the most prominent.}
\label{3D_kde}
\end{figure*}

Figure~\ref{3D_kde} presents the three-dimensional kernel density estimation (KDE) of burst energies, waiting times, and MJD, along with its two-dimensional marginal projections.
The plot reveals that burst activity is predominantly concentrated within a few active epochs, with MJD~60381 being the most prominent.
To robustly investigate short-timescale behavior, we selected the 3,197 bursts detected on MJD~60381 (spanning 4.38~hr) as a benchmark subset. This epoch is particularly well suited for this purpose, as it provides the longest continuous monitoring window and the highest burst rate, thereby enabling reliable analysis of rapid variations.

\subsection{Event-Rate Coherence Analysis}
To search for coherent growth in the burst rate, we adopt the method of \citet{2023ApJ...949L..33W}. For a given time resolution $\delta T$, we divide the observing window into contiguous bins of width $\delta T$, compute the rate $\lambda_i = N_i/\delta T$ for each bin, and identify sequences of at least four consecutive bins for which the rates are strictly increasing. For each detected structure, we fit the rate evolution as
\begin{equation}
\lambda(t) = \lambda(t_1) + (\lambda_0 - \lambda(t_1)) \left(\frac{t - t_1}{t_0 - t_1}\right)^p,
\label{eq:coherence}
\end{equation}
where $p>1$ indicates super-linear growth. We perform this search over a range of $\delta T$ values, from 20~s to 3800~s, using a sliding-window approach. This analysis is restricted
to the MJD~60381 subset, as it offers the longest contiguous interval and the highest event rate, which are essential for reliably detecting rate-coherence features.

\subsection{Waiting-Time Distribution Fitting}
The waiting time $\Delta t$ is defined as the interval between successive bursts. To construct the differential distribution $dN/dx$, we adopt logarithmic binning with $N_{\rm bins} = \mathrm{round}[5\log_{10}(x_{\max}/x_{\min})]$. For the $i$-th bin, the differential count and its associated uncertainty are computed as
\begin{equation}
\frac{dN}{dx}(x_i) = \frac{N_i}{\Delta x_i},\quad \sigma_i = \sqrt{N_i}/\Delta x_i,
\label{eq:diff_wtd}
\end{equation}
where $N_i$ is the number of events in the bin and $\Delta x_i$ is the bin width.
We fit the observed distribution with superpositions of exponential (Exp) and threshold power-law (TPL) components, defined respectively as
\begin{equation}
f_{\rm Exp}(x) = A \exp(-x/x_c),\quad f_{\rm TPL}(x) = B\,(x + x_0)^{-\alpha},
\label{eq:components}
\end{equation}
where $x_0$ represents a characteristic threshold offset. 
For the full sample, we compare three candidate models: two exponentials (2Exp), one exponential plus one TPL (Exp+TPL), and two exponentials plus one TPL (2Exp+TPL). For the MJD~60381 subset, we instead compare 2Exp, Exp+TPL, and three exponentials (3Exp). Model selection is performed using the Bayesian Information Criterion (BIC) and the reduced chi-square statistic ($\chi^2_\nu$), which jointly assess goodness-of-fit and penalise model complexity. All parameters reported in this work, including those from the waiting-time fits, and the rescaled range (R/S) analysis, and the energy distribution analysis described below, are estimated via MCMC using the \texttt{emcee} package \citep{2013PASP..125..306F} with uniform priors.

\subsection{Rescaled Range (R/S) Analysis}
Following \citet{2023ApJ...949L..33W}, we apply R/S analysis \citep{1969WRR.....5..967M} to the burst sequences ordered by burst index $n$.
For each lag $\tau_n$ (block size), we compute the averaged R/S as
\begin{equation}
R/S(\tau_n) = \frac{1}{\tau_n} \sum_{i=1}^{\lfloor N/\tau_n \rfloor} \frac{R_i(\tau_n)}{S_i(\tau_n)},
\label{eq:rs}
\end{equation}
where $R_i$ is the range of cumulative deviations from the mean within the $i$-th block, $S_i$ is the corresponding standard deviation, and the average is taken over all blocks. If the relation $R/S \propto \tau_n^{H}$ holds, the slope of the power law directly yields the Hurst exponent $H$. 
The Hurst exponent serves as a useful indicator of the presence and nature of memory in a time series. A value of $H = 0.5$ corresponds to a purely stochastic process with no long-range temporal correlations, whereas $H>0.5$ indicates persistence (positive memory) and $H<0.5$ indicates anti-persistence.
We sample the lags uniformly on a logarithmic grid, with $\tau_{\min}=10$ and $\tau_{\max}=N/2$. This analysis is performed separately for the waiting-time and energy sequences. For the MJD~60381 subset, we fit a single power law, whereas for the full sample we adopt a broken power-law model for both sequences.

\subsection{Energy Distribution Fitting}
For the energy distribution, we adopt the fitting methodology and functional forms of W.-L. Zhang et al. (in prep.). Based on their model comparison results, we adopt the Low-break Cutoff Power Law (Lb-CPL) as our primary model for the differential energy distribution:
\begin{equation}
\frac{dN}{dE} \propto \exp(-E_b/E)\, E^{-\alpha}\, \exp(-E/E_c),
\label{eq:LbCPL}
\end{equation}
where $E_b$ is the low-energy break parameter, $E_c$ the high-energy cutoff, and $\alpha$ the power-law index. We separately fit this model to bursts with waiting time $<0.25$~s (``short-wait'') and $\ge0.25$~s (``long-wait''), for both the full sample and the MJD~60381 subset. Best-fit parameters are obtained via MCMC following the same fitting procedure as in W.-L. Zhang et al. (in prep.), and the resulting $\alpha$ values are compared between the two waiting-time groups to investigate possible energy--waiting-time correlations.

\section{RESULTS}
\label{sec:results}
Table~\ref{tab_summary} summarizes the main results of our analyses.

\subsection{Event-Rate Coherence: The Longest Memory Timescale Detected to Date}
\label{subsec:coherence_results}
On MJD~60381, we detected 87 coherent structures across time resolutions $\delta T$ ranging from 20~s to 3800~s. A subset of the fitted nonlinearity indices satisfies $p\ge2.0$, indicating super-linear growth. Remarkably, we identified coherent structures up to $\delta T = 3600$~s, the longest such timescale ever reported for any repeating FRB, significantly exceeding the 2200~s reported for FRB~20121102A \citep{2023ApJ...949L..33W}. We note that the true memory timescale may be even longer, since our search is intrinsically limited by the 4.38-hr duration of the observing window on MJD~60381.

\begin{figure*}[http!]
\centering
\includegraphics[width=0.3\textwidth]{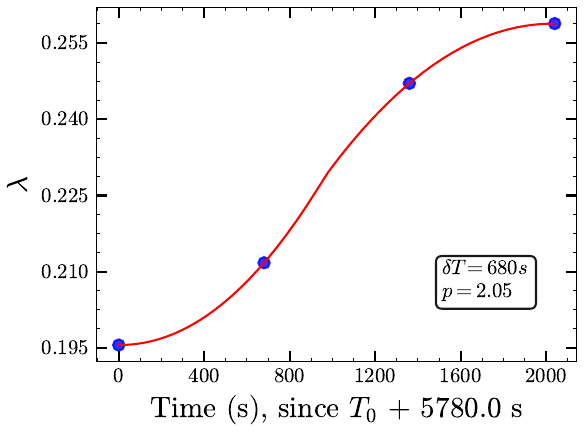}
\includegraphics[width=0.3\textwidth]{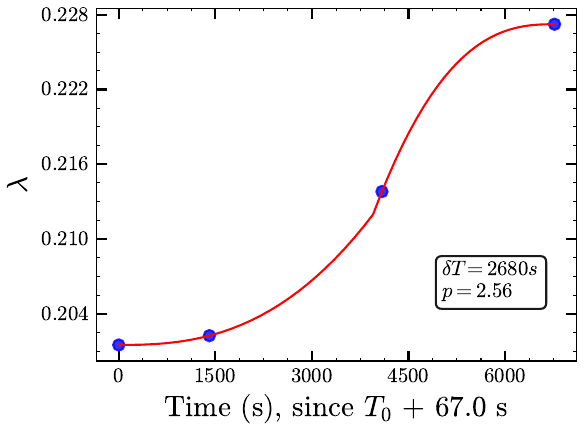}
\includegraphics[width=0.3\textwidth]{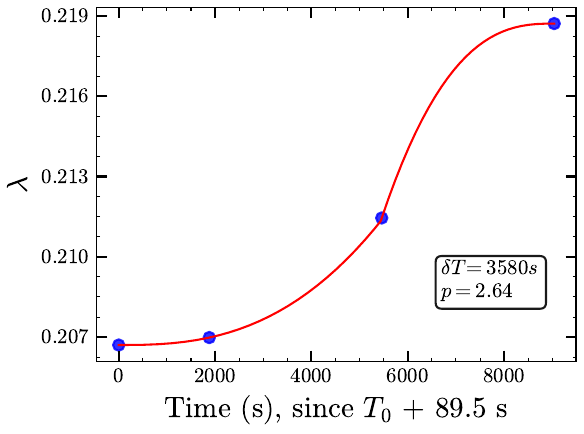}\\
\includegraphics[width=0.8\textwidth]{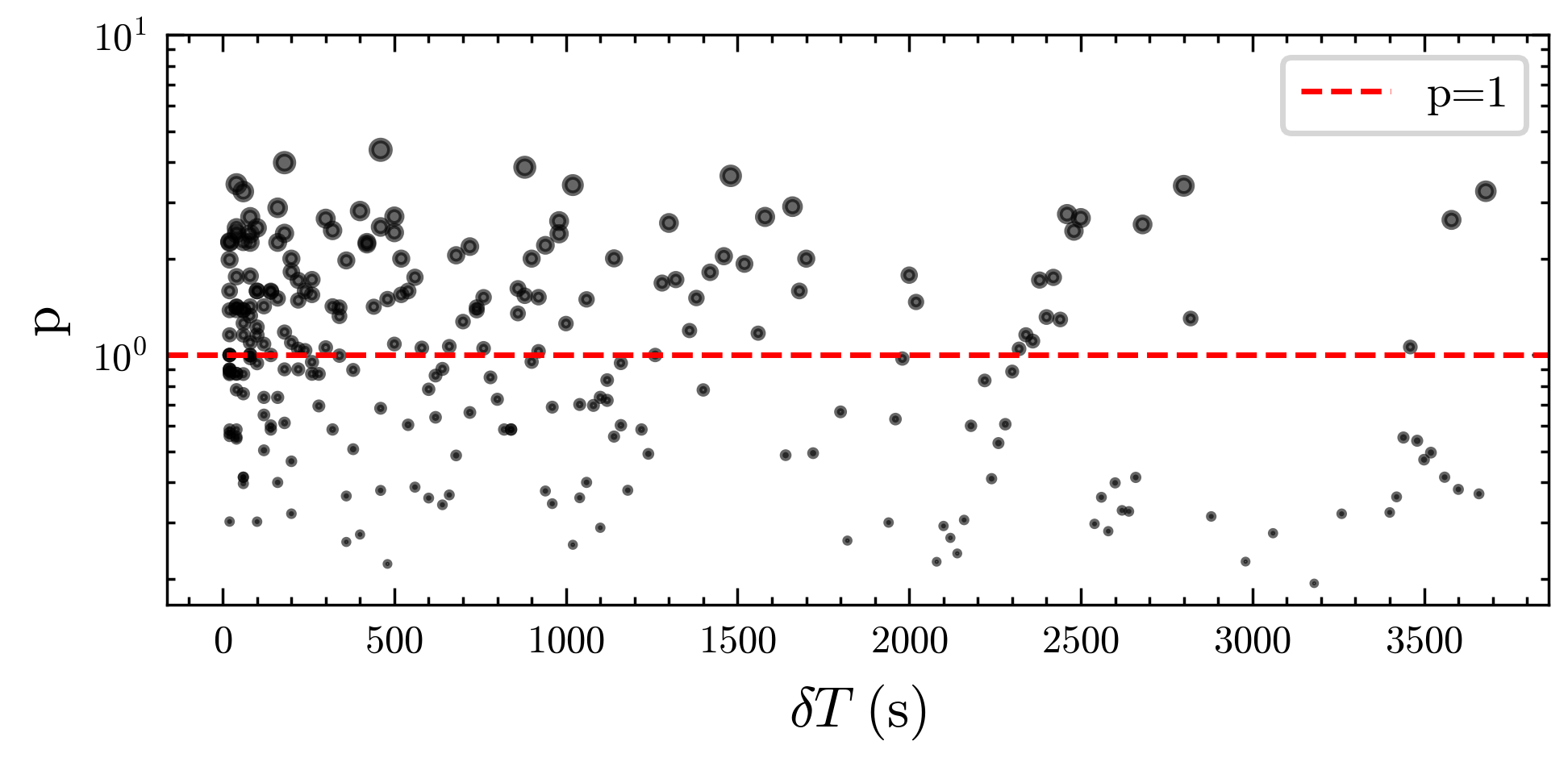}
\caption{Examples of coherent burst-rate structures detected on MJD~60381 at $\delta T = 680$~s, 2680~s, and 3580~s (top). Blue points show the observed rate; red curves are best-fit polynomials with the corresponding $p$ values indicated. Bottom: distribution of fitted $p$ values over all detected structures.}
\label{fig_coherence_example1}
\end{figure*}

Figure~\ref{fig_coherence_example1} shows representative coherent structures at $\delta T = 680$~s, 2680~s, and 3580~s. The presence of such coherent blocks implies that even when the waiting-time distribution is purely exponential (suggesting a Poisson process), the burst rate nonetheless contains a deterministic memory component operating on timescales from minutes to approximately one hour.

\subsection{Waiting-Time Distributions: From Exponentials to Power-Law Tails}
\label{subsec:wtd}
\begin{figure*}[http!]
\centering
\includegraphics[width=\columnwidth]{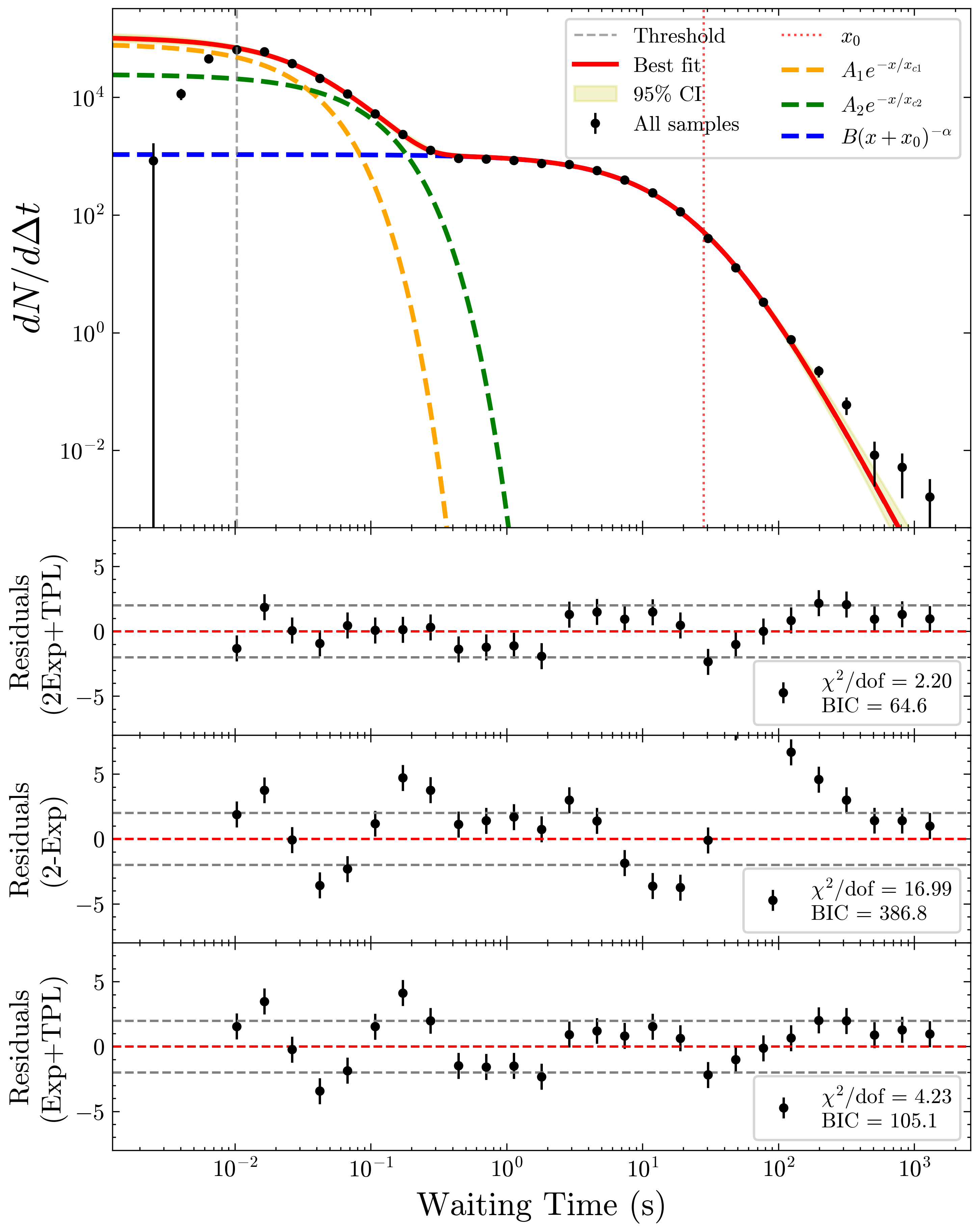}\includegraphics[width=\columnwidth]{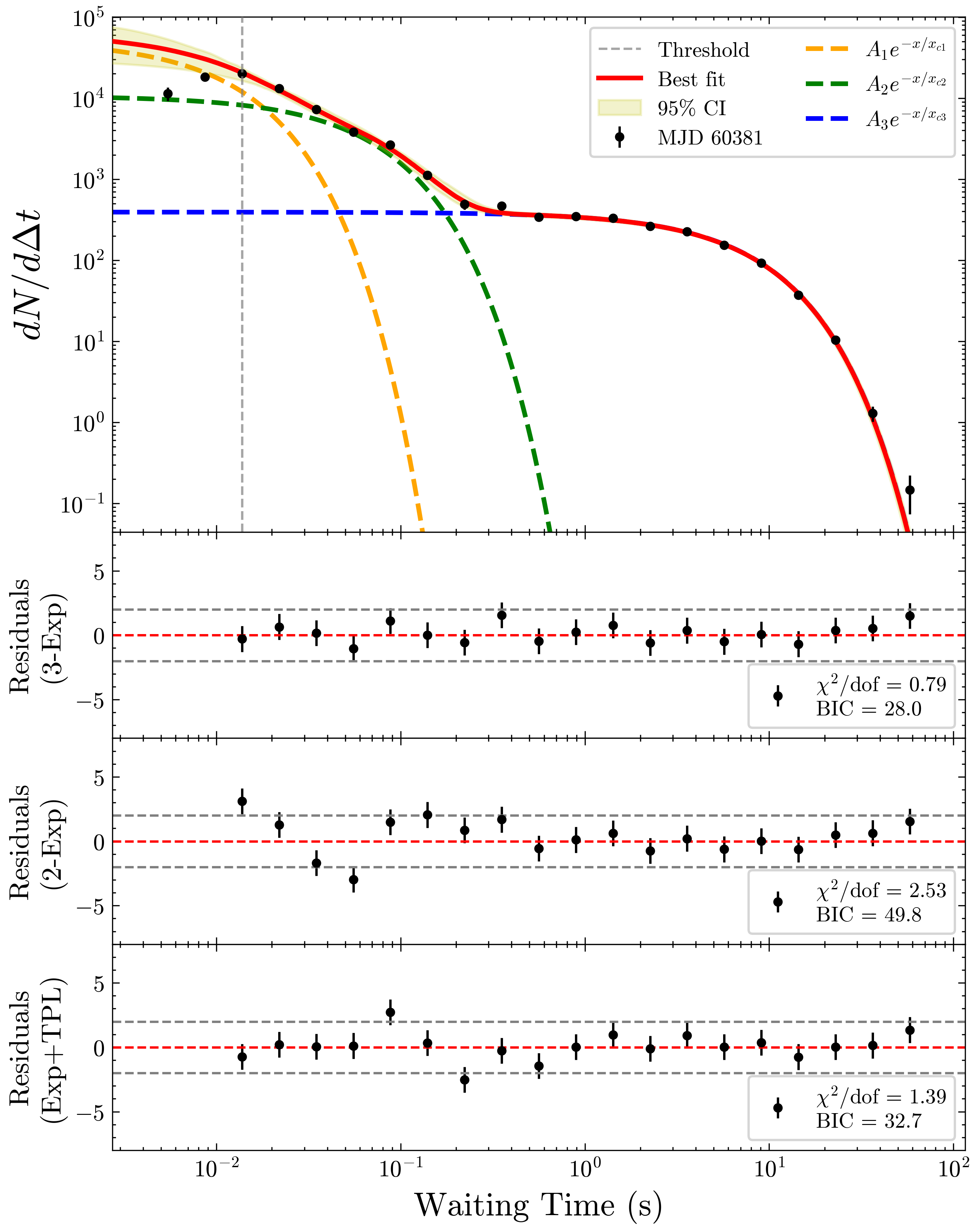}
\caption{Left (all-sample): best-fit 2Exp+TPL. Right (MJD~60381): best-fit 3Exp. The lower subpanels show residuals of competing models, with $\chi^2_\nu$ and BIC values indicated.}
\label{WTD_fit}
\end{figure*}

Figure~\ref{WTD_fit} presents the differential waiting-time distributions for the two data subsets. For the MJD~60381 subset (right panel), the distribution is well described by a superposition of three exponentials (3Exp), with no power-law tail required. The BIC strongly favors this 3Exp model over both the 2Exp and Exp+TPL alternatives ($\Delta\mathrm{BIC}>10$). This multi-exponential decomposition reveals three distinct characteristic timescales operating concurrently, a finding independently corroborated by L.-X. Zhang et al. (in prep.) using a different burst identification criterion.

In striking contrast,  the full-sample distribution (left panel) exhibits a qualitatively different structure. The short-waiting-time regimes ($\Delta t\lesssim0.3$~s) is fitted by 2Exp, while the long tail ($\Delta t\gtrsim30$~s) requires a threshold power law rather than a third exponential. Model selection again favours 2Exp+TPL over 3Exp ($\Delta\mathrm{BIC}>10$). The emergence of a power-law tail when the observational window expands from a single day to seven months is a hallmark of SOC systems, indicating that long-range temporal correlations only become apparent on longer timescales.

\subsection{R/S Analysis: From Random Walk to Non-Stationary Trend}
\label{subsec:rs}
\begin{figure*}[http!]
\centering
\includegraphics[width=\columnwidth]{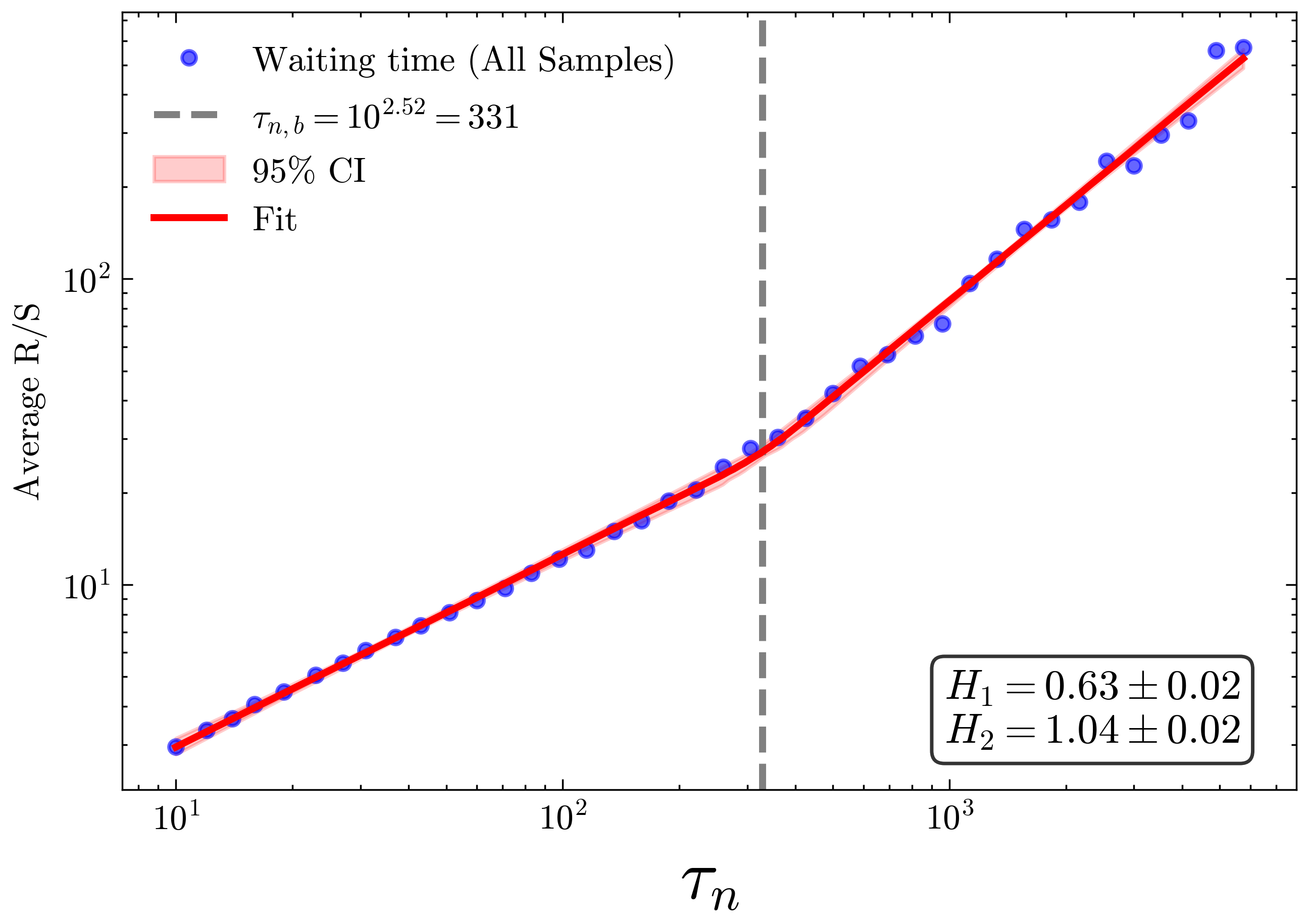}\includegraphics[width=\columnwidth]{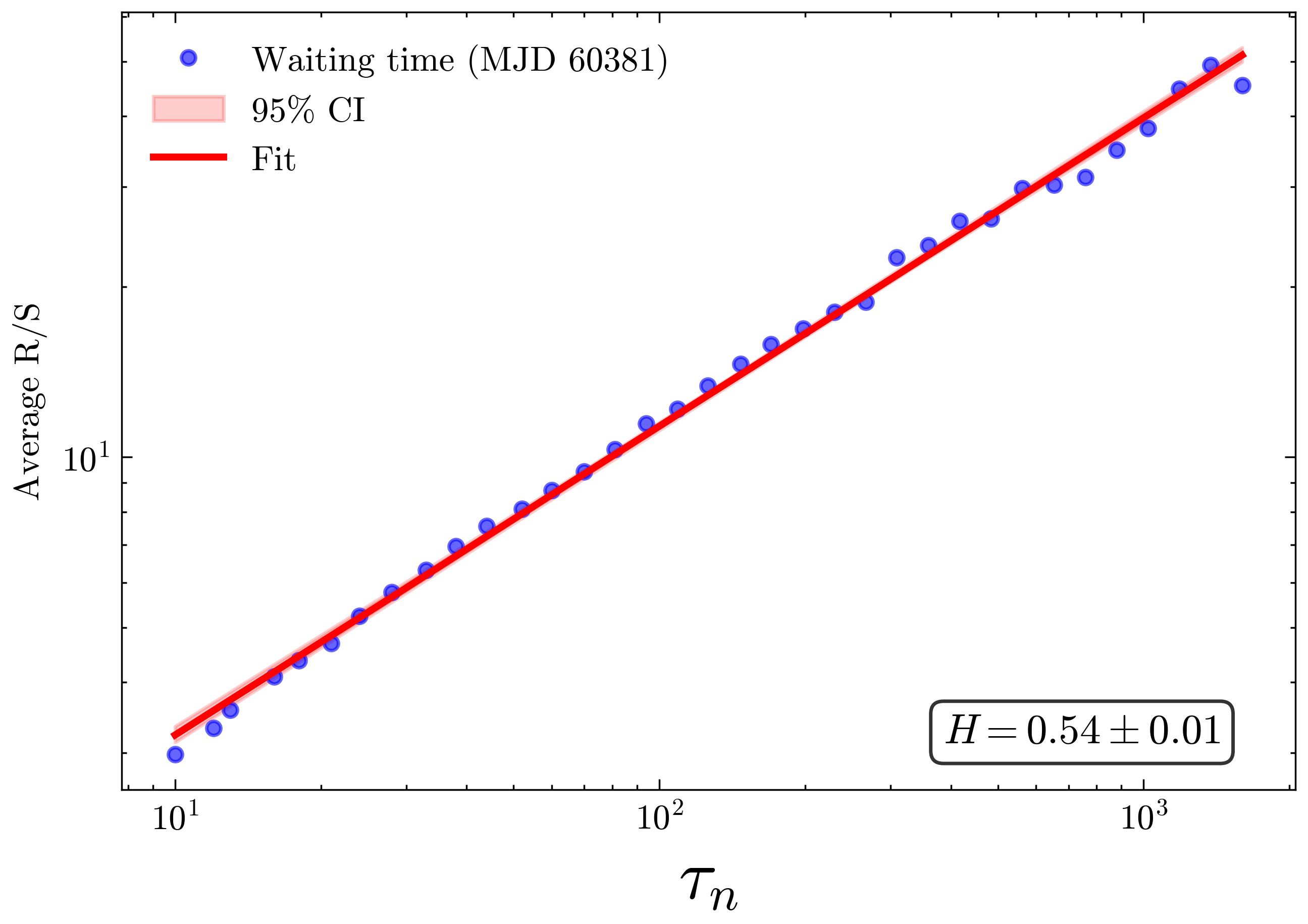}
\caption{Left: all-sample waiting-time R/S with broken power law ($H_1=0.63$, $H_2=1.04$, break at $\tau_{n,b}\simeq331$). Right: MJD~60381 with single power law ($H=0.54$).}
\label{fig_rs_waiting}
\end{figure*}

Figure~\ref{fig_rs_waiting} displays the R/S curves for waiting times. For MJD~60381 (right panel), a single power-law  fit yields $H=0.54\pm0.01$, consistent with a nearly random (Brownian) process. For the full sample (left panel), however, the R/S curve exhibits a clear break in slope. A broken power-law fit gives
\begin{align}
H_1 &= 0.63 \pm 0.02 \quad (\tau_n \le \tau_{n,b}),\\
H_2 &= 1.04 \pm 0.02 \quad (\tau_n > \tau_{n,b}),
\end{align}
with $\tau_{n,b}\approx331$. The short-lag exponent $H_1=0.63$ is only slightly above the random-walk value of 0.5, indicating weak persistence. The long-lag exponent $H_2=1.04>1$ indicates non-stationary behavior, a slow systemic drift over the 214-day observational baseline.

Converting the break lag $\tau_{n,b}$ to a physical timescale using the average burst rate ($\bar{R}\approx341~\mathrm{hr^{-1}}$) gives $\tau_{n,b}/\bar{R}\approx3495~\mathrm{s}\approx58$~min, which is remarkably close to the 3600~s coherence limit identified in Section~\ref{subsec:coherence_results}. Although this conversion is approximate given the non-uniform burst rate, the proximity of these two independent estimates suggests that they may trace a common physical scale.

\begin{figure*}[http!]
\centering
\includegraphics[width=\columnwidth]{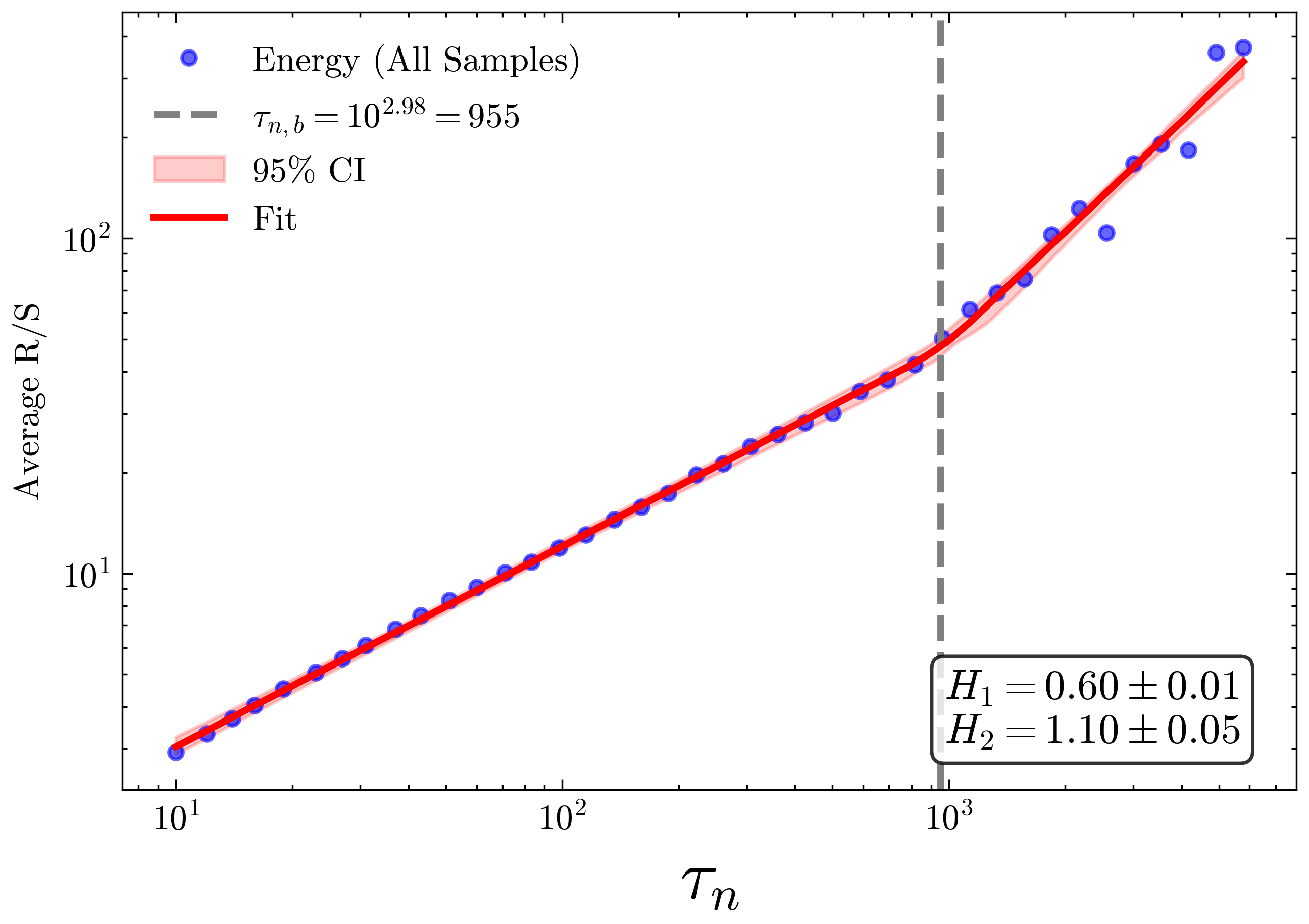}\includegraphics[width=\columnwidth]{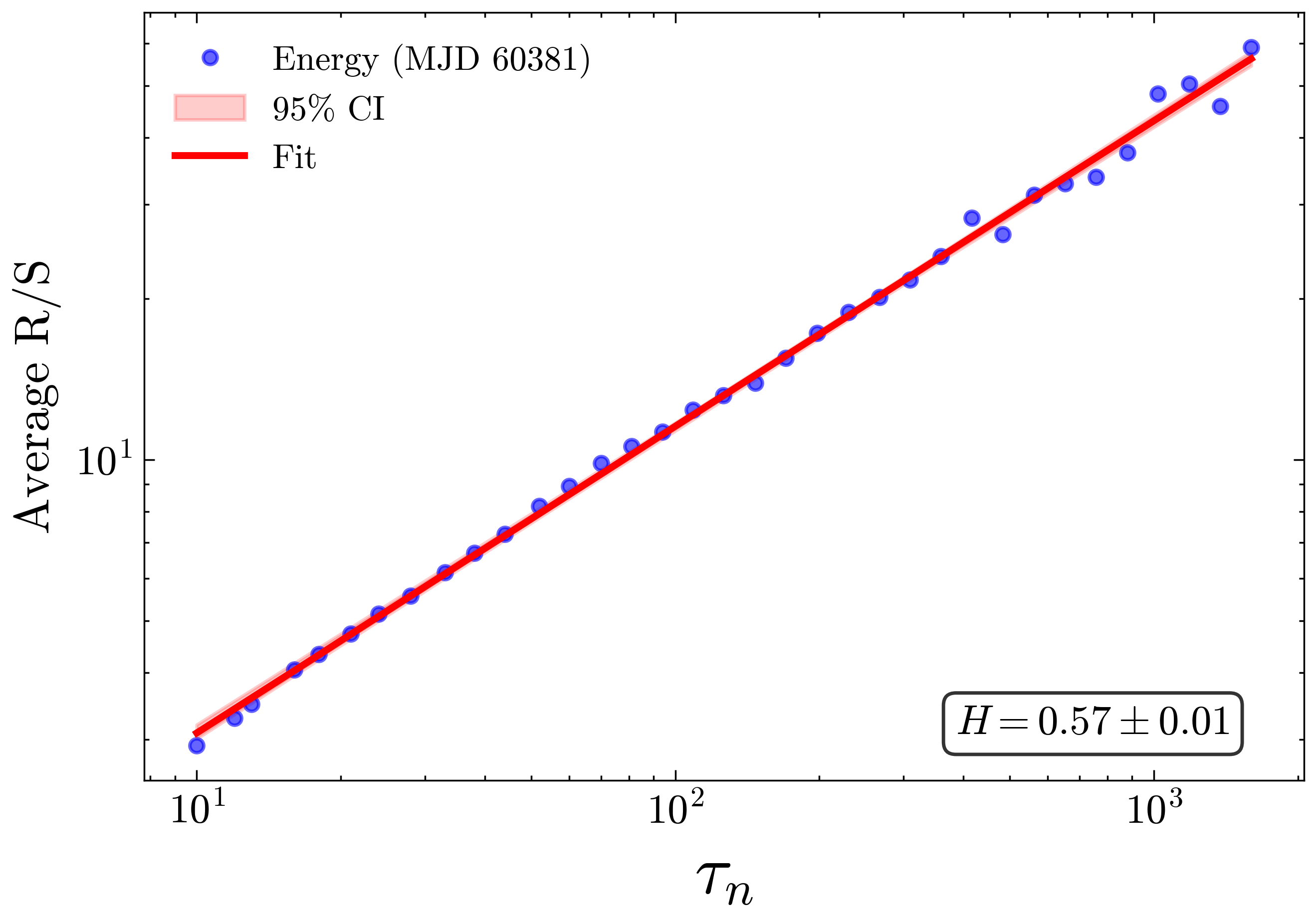}
\caption{Left: all-sample energy R/S with broken power law ($H_1=0.60$, $H_2=1.10$, break at $\tau_{n,b}\simeq955$). Right: MJD~60381 energy R/S with single power law ($H=0.57$).}
\label{fig_rs_energy}
\end{figure*}

Figure~\ref{fig_rs_energy} shows the corresponding R/S curves for the energy sequences. For MJD~60381, we obtain $H=0.57\pm0.01$, again consistent with near-random behavior. For the full sample, the energy R/S curve also exhibits a clear break, with a broken power-law fit yielding
\begin{align}
H_1 &= 0.60 \pm 0.01 \quad (\tau_n \le \tau_{n,b}),\\
H_2 &= 1.10 \pm 0.05 \quad (\tau_n > \tau_{n,b}),
\end{align}
and $\tau_{n,b}\approx955$. The short-lag exponent $H_1=0.60$ again suggests weak memory close to random behavior, while the presence of a break and $H_2>1$ indicates that the energy sequence also exhibits non-stationary behavior on long lags. This is consistent with the picture of a slow systemic change over the full observational span. The break lag for the energy sequence differs from that of the waiting-time sequence, which may reflect that the two sequences probe different aspects of the burst process (i.e., temporal clustering versus energy release) and are thus sensitive to different physical scales.

\subsection{Energy Distributions: Waiting-Time-Dependent Gutenberg-Richter Indices}
\label{subsec:energy}
\begin{figure*}[http!]
\centering
\includegraphics[width=\columnwidth]{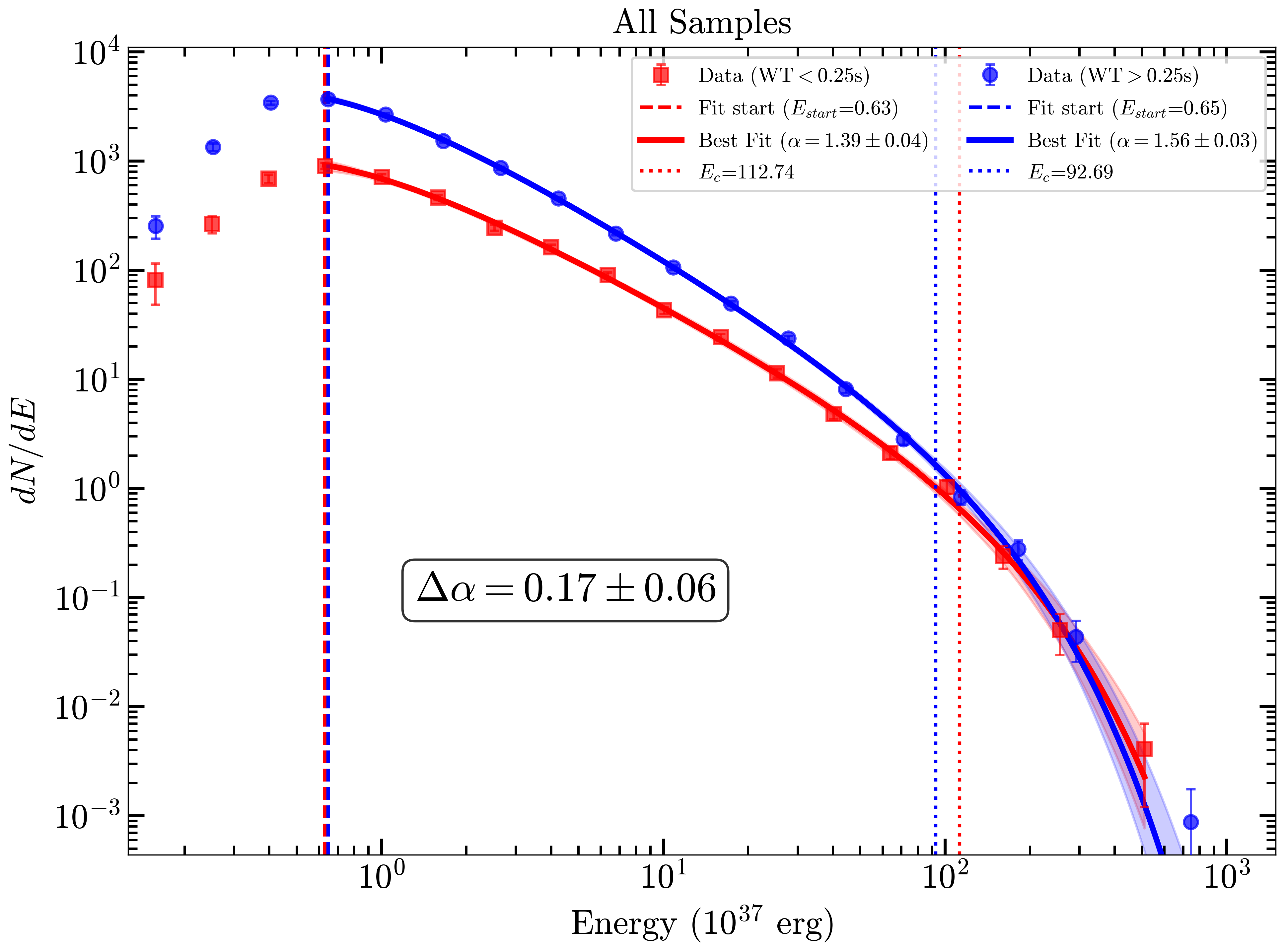}\includegraphics[width=\columnwidth]{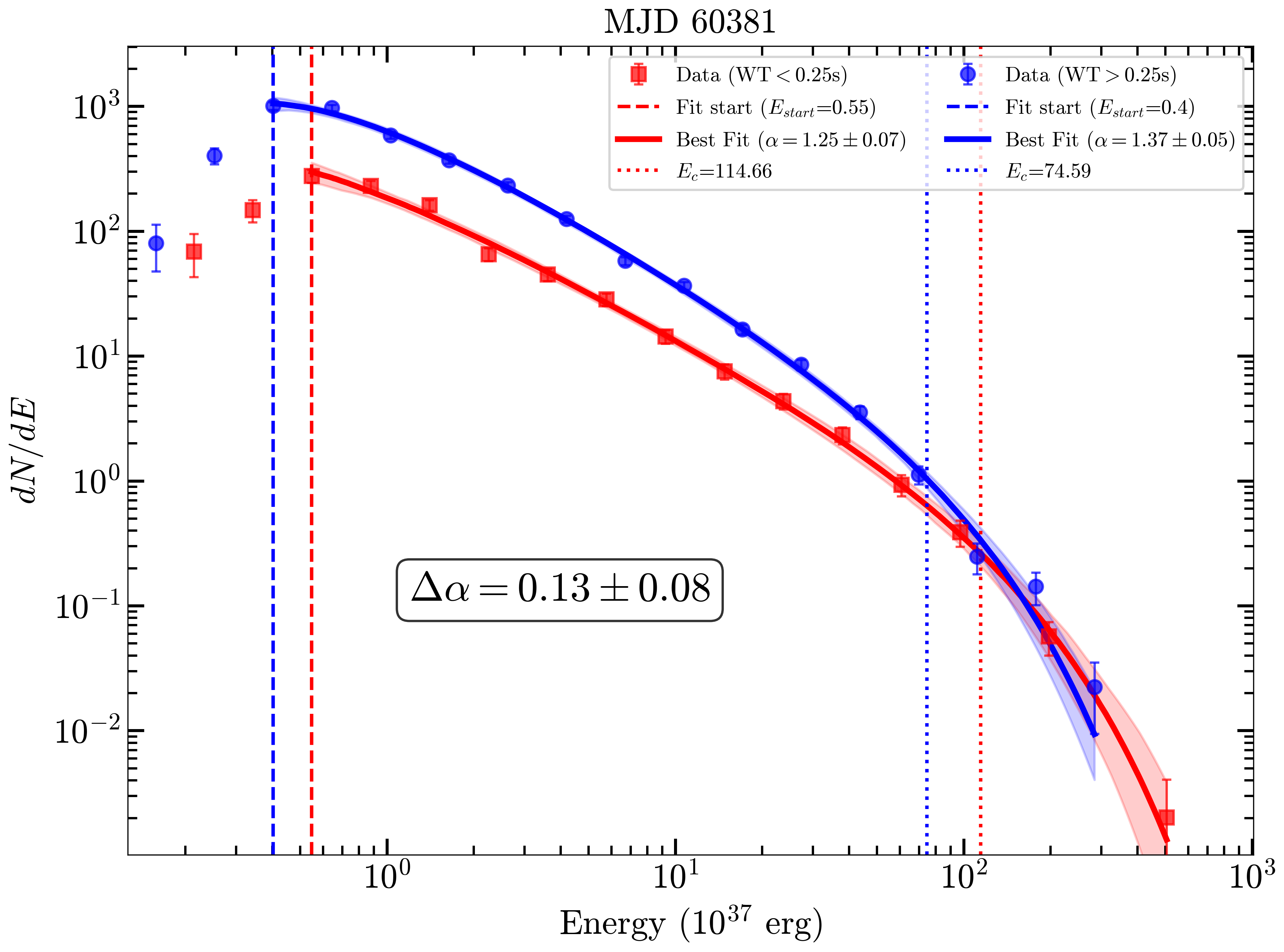}
\caption{Differential energy distributions for short-wait ($<0.25$~s, red) and long-wait ($\ge0.25$~s, blue) bursts. Left: full sample. Right: MJD~60381. Solid curves show the best-fit Lb-CPL models, with the power-law indices indicated.}
\label{fig_energy_split}
\end{figure*}

Figure~\ref{fig_energy_split} shows the differential energy distributions for short-wait and long-wait bursts. For the full sample, the Lb-CPL fits yield
\begin{equation}
\alpha_{\mathrm{short}} = 1.39 \pm 0.04,\qquad \alpha_{\mathrm{long}} = 1.56 \pm 0.03,
\end{equation}
giving a difference of $\Delta\alpha = 0.17 \pm 0.05$. For the MJD~60381 subset, we obtain
\begin{equation}
\alpha_{\mathrm{short}} = 1.25 \pm 0.07,\qquad \alpha_{\mathrm{long}} = 1.37 \pm 0.05,
\end{equation}
with $\Delta\alpha = 0.13 \pm 0.09$. These $\Delta\alpha$ values are consistent between the two samples within uncertainties, suggesting a scale-invariant feature. Physically, the lower $\alpha$ (shallower index) during rapid bursting parallels the Gutenberg-Richter $b$-value behavior in earthquakes, where $b$ decreases with increasing shear stress. The high-energy cutoff $E_c$ remains consistent between short-wait and long-wait bursts, suggesting the maximum energy scale is an intrinsic source property.

\begin{table*}
\centering
\caption{Summary of statistical properties for FRB~20240114A}
\label{tab_summary}
\begin{tabular}{l c c}
\hline
Property & All-sample (214 days) & MJD~60381 (4.38 hr) \\
\hline
Event-rate coherence max $\delta T$ & Not performed & $\sim 3600$~s \\
Waiting-time distribution & 2Exp+TPL & 3Exp \\
Hurst exponent (waiting time) & Broken: $H_1=0.63\pm0.02$, $H_2=1.04\pm0.02$ & $H=0.54\pm0.01$ \\
Hurst exponent (energy) & Broken: $H_1=0.60\pm0.01$, $H_2=1.10\pm0.05$ & $H=0.57\pm0.01$ \\
R/S break $\tau_{n,b}$ (average timescale) & 331($\sim 1$~hr) for $\Delta t$, 955($\sim 3$~hr) for $E$ & N/A \\
Energy index ($\alpha$) & $\alpha_{\text{short}}=1.39\pm0.04$, $\alpha_{\text{long}}=1.56\pm0.03$ & $\alpha_{\text{short}}=1.25\pm0.07$, $\alpha_{\text{long}}=1.37\pm0.05$ \\
$\Delta\alpha = \alpha_{\text{long}}-\alpha_{\text{short}}$ & $0.17\pm0.05$ & $0.13\pm0.09$ \\
\hline
\end{tabular}
\end{table*}

\section{DISCUSSION}
\label{sec:discussion}

\subsection{The Four Pillars of Evidence for Scale-Dependent Memory}

Our multi-pronged analysis yields four independent lines of evidence that collectively point to scale-dependent memory in FRB~20240114A.

\textbf{1. The longest coherent event-rate growth.} On MJD~60381, we detected coherent event-rate structures up to 3600~s, the longest timescale reported for any repeating FRB to date. This finding indicates that even when the waiting-time distribution is purely exponential, the burst rate retains a deterministic memory component operating on timescales up to approximately one hour.

\textbf{2. The exponential-to-power-law transition in waiting-time distributions.} The MJD~60381 distribution is described by three exponentials, while the full 214-day sample requires a power-law tail. This transition is a hallmark of SOC systems, indicating that long-range temporal 
correlations only become manifest over sufficiently long timescales. Intriguingly, the three-exponential form observed on the single day may hint at the co-existence of multiple underlying physical processes operating concurrently.

\textbf{3. The R/S analysis of the full sample reveals broken power laws for both waiting-time and energy sequences.} For waiting times, we find $H_1=0.63$, near-random with weak persistence, and $H_2=1.04$, indicative of non-stationary behavior. The energy sequence yields a qualitatively similar pattern, with $H_1=0.60$ and $H_2=1.10$. The fact that $H_2>1$ in both cases points to a slow systemic drift that is not captured by a stationary stochastic model.

\textbf{4. The energy distribution indices exhibit a systematic dependence on waiting time, with $\alpha_{\rm short}<\alpha_{\rm long}$ and a consistent $\Delta\alpha$ across both the full sample and the daily subset.}
The energy distribution indices exhibit a systematic dependence on waiting time, with $\alpha_{\rm short}<\alpha_{\rm long}$ and a consistent $\Delta\alpha$ across both the full sample and the daily subset. This suggests a scale-invariant stress-dependent behavior that parallels the Gutenberg-Richter relation in seismology.

Taken together, these four pillars---coherent rate growth, regime transition in waiting times, broken R/S scaling, and waiting-time-dependent energy indices---provide robust, mutually reinforcing evidence for memory that operates differently on short and long timescales.

\subsection{Physical Interpretation of the Three-Component Exponential Decomposition}

The three-exponential decomposition of the MJD~60381 waiting-time distribution reveals multiple characteristic timescales operating concurrently. Unlike a single exponential, which would imply a homogeneous Poisson process, this multi-exponential structure suggests a superposition of multiple Poisson-like processes with distinct rate parameters. This finding is independently corroborated by L.-X. Zhang et al. (in prep.), who employed a different burst identification criterion, lending confidence to its robustness.
It is tempting to speculate that these components correspond to several independent triggering channels operating simultaneously; their superposition would naturally yield a multi-exponential waiting-time distribution. A slow modulation of their relative weights, possibly by a gradually evolving global magnetospheric parameter, could then account for the event-rate coherence on longer timescales, providing a potential bridge between the short-term exponential statistics and the observed long-term memory.

\subsection{The Regime Shift: From Local Randomness to Global Non-Stationarity}

Our results indicate that FRB~20240114A does not occupy a single dynamical state. On short timescales (e.g., the 4.38~hr window of MJD~60381), the source behaves as a nearly stationary random process: the waiting-time distribution is purely exponential, and the R/S analysis yields $H\approx0.54$ for waiting times and $H\approx0.57$ for energies, both consistent with a stochastic process lacking long-range memory. On the 214-day baseline, however, we observe systematic drifts in both the waiting-time and energy sequences, with $H_2>1$ for both, accompanied by the emergence of a power-law tail in the waiting-time distribution.

This regime shift is naturally explained if the magnetar's burst rate is modulated by a slowly evolving parameter, such as global magnetic stress, spin-down torque, or magnetospheric twist angle. The exponent $H_2>1$ is a hallmark of non-stationary fractional Brownian motion, where the variance grows faster than linearly with time. The waiting-time R/S break at $\sim$1 hour remarkably coincides with the coherence detection limit ($\delta T = 3600$~s) from the daily analysis, suggesting that these independent diagnostics may trace a common timescale. The energy R/S break occurs at a different lag, which is not surprising given that waiting times and energies probe different physical aspects of the burst process---the former reflects triggering dynamics, while the latter reflects energy release. Whether these timescales are intrinsic to the magnetar or reflect the observational window remains an open question, motivating future long-duration monitoring of this and other hyperactive repeaters.

It is instructive to compare our findings with the recent work of \citet{Xu2026}, who used the Pincus Index and Lyapunov Exponent to characterize repeating FRBs as highly stochastic and weakly chaotic on average. Their global diagnostics place FRBs in a distinct region of the stochasticity-chaos phase space, separate from magnetar flares and pulsar glitches. Our scale-dependent R/S analysis complements this picture by revealing that the apparent stochasticity is not uniform across all timescales. While the source indeed appears nearly random on short timescales, consistent with their global characterization, the long-term evolution exhibits systematic non-stationarity ($H_2>1$) that is not captured by global entropy-based measures. The two perspectives are therefore complementary: \citet{Xu2026} provide the global portrait, while our multi-scale analysis uncovers the structural details at different temporal resolutions.

Notably, the R/S analyses of waiting times and energies yield qualitatively consistent results: both exhibit a break and $H_2>1$ on long lags, indicating non-stationary behavior across two independent sequence properties—temporal clustering and energy release. This consistency, while not definitive, lends additional support to the robustness of the observed regime shift and suggests that the non-stationarity is not merely a feature of a single statistic.

The multi-scale framework established here also opens the door for comparative studies across different repeaters. It will be important to determine whether similar scale-dependent transitions exist in other sources with sufficient burst statistics, and whether the timescale of the break correlates with other source properties such as rotation period or magnetic field strength.

\subsection{Energy Distribution and Observational Selection Effects}

The observed $\Delta\alpha$ between short-wait and long-wait bursts is consistent across the all-sample and the daily subset, suggesting that this is an intrinsic physical signature rather than a purely observational artifact. A lower $\alpha$ (shallower index) during rapid bursting episodes parallels the behavior of the Gutenberg-Richter $b$-value in earthquakes\citep{1956BuSSA..46..105G}, which decreases with increasing shear stress. This suggests that during high-rate bursting episodes, the system is in a higher stress state where large events are relatively more frequent. The consistency of the high-energy cutoff $E_c$ between short-wait and long-wait bursts for both samples further supports the view that the maximum energy scale is an intrinsic source property.

However, we caution that observational selection effects may contribute to the measured difference. Bursts with shorter waiting times occur in denser sequences, and lower-energy bursts within such sequences are more likely to be missed due to pulse overlap or limited signal-to-noise ratio. This would artificially steepen the energy distribution for short-wait bursts, making $\alpha_{\rm short}$ appear larger and reducing the true $\Delta\alpha$. Conversely, long-wait bursts are more isolated and may suffer less from such incompleteness. The ratio of the two power-law distributions scales as $E^{-\Delta\alpha}$, which could serve as a first-order quantification of such observational biases.

A more subtle possibility is that a nearly uniform observational bias operates across all samples, artificially preserving the invariance of $\Delta\alpha$. However, such a bias would need to remain effective across burst rates spanning two orders of magnitude while leaving the high-energy cutoff $E_c$ unaffected. The fact that $\Delta\alpha$ does not increase on the densest day (MJD~60381), where pulse-overlap incompleteness would be most severe, argues against a purely selection-driven origin. If the true $\Delta\alpha$ is smaller than the measured value, then the stress-dependent SOC interpretation would be weakened, and the waiting-time-dependent energy distribution might be predominantly a detection effect. Nevertheless, quantitative disentanglement of intrinsic and observational contributions ultimately requires future observations with higher sensitivity and better time resolution.

\subsection{Comparison with Previous Work}

Our results significantly extend those of \citet{2023ApJ...949L..33W}. While they reported $H=0.62$ and $0.70$ for FRB~20121102A and FRB~20201124A, respectively, indicative of stationary memory, our detection of $H_2>1$ reveals a non-stationary regime not accessible with shorter observational baselines. The longer coherence timescale ($\ge3600$~s versus 2200~s) suggests that FRB~20240114A may possess a more extended energy reservoir or a slower triggering mechanism than previously studied repeaters.

Recent works have quantified randomness and scale invariance in repeating FRBs \citep{Sang2024} and uncovered memory from minutes to an hour \citep{2023ApJ...949L..33W}. Our findings are in broad agreement with these studies while revealing a new regime shift. The distinct location of FRBs in the stochasticity-chaos phase space \citep{Xu2026} supports the view that repeating FRBs are not simple analogues of magnetar flares or starquakes. Instead, they represent a unique class of high-entropy, weakly chaotic systems, a pocture consistent with our finding of near-random short-term behavior coexisting with long-term non-stationarity.

The energy index $\alpha\sim1.39$-$1.56$ is consistent with the FD-SOC prediction of $\alpha\sim1.5$ for three-dimensional energy dissipation \citep{2012A&A...539A...2A}. Moreover, \citet{2023ApJ...949L..33W} showed that power-law tails in waiting-time distributions can be reproduced by a non-stationary Poisson process driven by correlated external perturbations. 
Our R/S results independently confirm the presence of such non-stationarity, lending empirical support to this theoretical picture.

\subsection{Limitations and Caveats}

Several caveats should be borne in mind when interpreting our results. First, the R/S analysis treats waiting times as a sequence indexed by burst order rather than physical time; the inferred break timescale therefore depends on converting burst counts to temporal units via the average burst rate. Second, this conversion is necessarily approximate given the highly variable burst rate, and the close correspondence between the R/S break and the coherence limit should be interpreted with appropriate caution. Third, the energy distribution fits are subject to incompleteness at low energies, as discussed above, which may systematically affect the derived power-law indices. Fourth, the coherence analysis is intrinsically limited by the 4.38~hr observing window on MJD~60381; the true memory timescale may therefore exceed 3600~s, and our reported value should be regarded as a lower limit. Fifth, our analysis is based on a single source; generalisation to the broader FRB population will require similar multi-scale studies of other repeaters with rich burst statistics.

Despite these limitations, the consistency of our results across multiple independent methods—coherence analysis, waiting-time fitting, R/S analysis, and energy distribution modelling—strengthens the overall conclusion that FRB~20240114A exhibits a genuine scale-dependent memory, with near-random behaviour on short timescales giving way to systematic non-stationarity over longer baselines.


\section{CONCLUSIONS}
\label{sec:conclusions}

We have performed a comprehensive statistical analysis of FRB~20240114A based on 11,553 bursts detected with FAST. Our results yield four indeoendent lines of evidence for scale-dependent memory in this source:

\begin{enumerate}
\item \textbf{The longest coherent event-rate growth detected to date.} On MJD~60381, we identified coherent rate structures extending up to 3600~s, the longest such timesacle ever reported for any repeating FRB. This finding demonstrates that memory persists on timescales of up to an hour, even when the waiting-time distribution is purely exponential.

\item \textbf{A transition from exponential to power-law waiting-time distributions.} The MJD~60381 subset is well described by a superposition of three exponentials (3Exp), a resulte independently confirmed by L.-X. Zhang et al. (in prep.) using a different burst identification criterion. In contrast, the full 214-day sample requires a threshold power-law tail (2Exp+TPL), marking a qualitative transition from Poisson-like statistics on short baselines to SOC-like behavior as the observational window expands.

\item \textbf{Broken power-laws in R/S analysis.} For the full sample, the waiting-time R/S curve exhibits a clear break, with $H_1=0.63\pm0.02$ (near-random on short lags) and $H_2=1.04\pm0.02$. The break occurs at a timescale of $\sim$1~hour, which closely matches the 3600~s coherence limit identified in the daily analysis. The energy R/S curve similarly shows a broken power law, with $H_1=0.60\pm0.01$ and $H_2=1.10\pm0.05$, though the break occurs at a different lag. The fact that $H_2>1$ in both cases reinforces the interpretation of non-stationary behavior across multiple sequence properties.

\item \textbf{Scale-invariant waiting-time-dependent energy distributions.} The energy distribution exhibits Gutenberg--Richter-like indices, with $\Delta\alpha \equiv \alpha_{\rm long}-\alpha_{\rm short}=0.17\pm0.05$ for the full sample and $0.13\pm0.09$ for the daily subset, consistent within uncertainties. The high-energy cutoff $E_c$ remains constant between short-wait and long-wait bursts, suggesting that the maximum energy scale is an intrinsic property of the source, independent of the burst rate.
\end{enumerate}

Taken together, these results demonstrate that FRB~20240114A undergoes a regime shift: locally, on short timescales (seconds to hours), the burst process appears nearly random and Poisson-like; globally, on longer timescales (weeks to months), it exhibits systematic evolution and non-stationarity. The $\sim$1~hour timescale emerges as a potentially fundamental scale of the burst dynamics, although its physical origin remains to be determined.

This multi-scale memory framework provides new observational benchmarks for burst models, illustrating that a single source can exhibit fundamentally different statistical behaviors depending on the observational timescale. These findings underscore the critical importance of long-term, high-cadence monitoring programmes, which are essential for uncovering hidden non-stationarity and for developing a comprehensive understanding of burst dynamics across the broader population of repeating FRBs.

\begin{acknowledgments}
We thank the FAST team for providing the high-quality burst catalog. This work is supported by the National Natural Science Foundation of China (grant Nos. 12422307, 12373053, and 12321003).
\end{acknowledgments}

\bibliography{sample701}{}
\bibliographystyle{aasjournalv7}

\end{CJK*}
\end{document}